\documentclass{article} 

\usepackage[dvips]{graphicx}

\title{Microlens Mass Functions} 
\author{William D. Heacox}         

\begin{document}

\maketitle 

\begin{abstract}

A non-parametric statistical model is constructed to directly relate the distribution of observed
microlens timescales to that of the mass function of the population from which the lenses are
drawn, corrected for observational selection based on timescales and event amplifications.  Explicit
distributions are derived for microlensing impact parameters and maximum amplifications; both are shown
to be statistically independent of all other parameters in the problem, including lens mass.  The model
is used to demonstrate that the narrow range of microlens timescales observed toward the Large Magellanic
Cloud (LMC) is probably not consistent with lensing by a widely distributed spheroidal population (``dark halo") of 
large velocity dispersion, but is consistent with lensing within a rotating thick disk.  Poor numerical
conditioning of the statistical connection between lens masses and event timescales, and small number
statistics, severely limit the mass function information obtainable from current microlensing surveys
toward the LMC.

\end{abstract}

\section{INTRODUCTION}

Microlensing is a promising tool for eliciting the properties of massive
compact objects in the Galactic halo and toward the Galactic bulge, and
several such observing programs are underway or have been completed.  In
particular, the MACHO and EROS projects have published results for several
years of monitoring $>10^{6}$ stars in the Large Magellanic Cloud (LMC;
Alcock et al. 2000, Milsztajn et al. 2001), which detected $10-20$ events; and
both the MACHO and OGLE projects (Udalski et al. 2000) have published the
results of microlensing surveys toward the Galactic bulge that include more
than $200$ events.  Several additional microlensing surveys are currently
underway.  The results of these surveys have provided otherwise
inaccessible information on the structure and contents of the Galaxy, but
not always in unambiguous ways.  The primary object of interest has been
the total mass in the form of compact objects, particularly in the
(presumably dark) Galactic halo, usually analyzed in the form of
microlensing optical depths originally proposed by Paczy\'{n}ski (1986). 
As useful as these analyses have been, little information on the mass
function of the population underlying the microlensing observations has been
forthcoming, although some parametric modeling (e.g., Green 2000) has
sufficed to indicate some possible constraints conditional upon various
kinematic models of the halo.  Here we propose a more direct way to
elucidate the mass function of the population responsible for microlenses,
in the form of a non-parametric statistical model that directly relates the
mass function to the microlens quantities being observed.  In principle,
such a model should allow an unambiguous determination of the mass function,
given only the kinematics of the population.  In practice, we expect that
the information obtainable from such a method will be limited by small
number statistics and ambiguities concerning the kinematics of the lensing
population.

The necessary microlens physical relations are as follows (see, e.g., Paczy\'{n}ski 
1996 or Gould 1996 for more complete discussions).  The physical
scale of a microlensing event in the lens plane (i.e., the plane of the sky
at the lens location) is given by the Einstein radius:
\begin{equation}
r_{E}=\frac{2}{c}\sqrt{\frac{GMD_{L}\left( D_{S}-D_{L}\right) }{D_{S}}}\;,
\label{RE}
\end{equation}
where $D_{L}$ and $D_{S}$ are, respectively, lens and source star distances
from the observer; and $M$ is the lens mass.  For solar-mass halo
microlenses viewed against the backdrop of the LMC, $r_{E}$ is typically a
few astronomical units.  In terms of this quantity, the scaled impact
parameter $\beta $ and event timescale $\mathcal{T}$ are defined to be%
\begin{eqnarray}
\beta &=&b/r_{E}\;,  \label{beta} \\
\mathcal{T} &=&r_{E}/V_{\perp }\;,  \label{tau}
\end{eqnarray}
where $b$, the \textquotedblleft impact parameter\textquotedblright , is the
closest approach to the lens of the line-of-sight to the source star and 
$V_{\perp }$ is the relative transverse velocity of the encounter, both
measured in the lens plane.  The maximum optical amplification of the event
is
\begin{equation}
\mathcal{A}=\frac{\beta ^{2}+2}{\beta \sqrt{\beta ^{2}+4}}\;\Rightarrow
\;\beta \left( \mathcal{A}\right) =\left[ 2\left( \frac{\mathcal{A}}{\sqrt{%
\mathcal{A}^{2}-1}}-1\right) \right] ^{1/2}\;.  \label{alpha}
\end{equation}
The quantities $\mathcal{A}$ and $\mathcal{T}$ are easily derived from the
event light curve, and constitute the physically significant observables for
simple (i.e., point-source and point-lens) microlenses.  

Within a homogeneous set of observed microlenses, the overall statistical
distribution of observables, $f_{\mathcal{A},\mathcal{T}}\left( \mathcal{A},
\mathcal{T}\right) $, can be formally related to that of lens mass among the
population from which they were drawn, $f_{M}\left( M\right) $, by a
relation of the form
\begin{equation}
f_{\mathcal{A},\mathcal{T}}\left( \mathcal{A},\mathcal{T}\right) =\int 
f_{\mathcal{A},\mathcal{T}|M}\left( \mathcal{A},\mathcal{T}|M\right) \,
f_{M}\left( M\right) \,dM\;.  \label{M1}
\end{equation}
The notation is that $f_{x}\left( y\right) $ is the probability density
function (pdf) of random variable $x$ evaluated at $x=y$; $f_{x|z}$ is the
pdf of $x$ conditional upon $z$.  The core of the model is the kernel
function $f_{\mathcal{A},\mathcal{T}|M}\left( \mathcal{A},\mathcal{T}
|M\right) $, the probability that an object of mass $M$ will yield a
microlensing event characterized by $\left( \mathcal{A},\mathcal{T}\right) $
.  This is essentially an accounting mechanism, one that adds up the
probabilities of all possible combinations of lensing parameters $\left(
b,V_{\perp },D_{L},D_{S}\right) $ that lead to the observed $\left( \mathcal{
A},\mathcal{T}\right) $, given mass $M$.  Its computation requires the
provision of a kinematic model of the lens and source populations; i.e., 
\textit{a priori} distributions of $\left( b,V_{\perp },D_{L},D_{S}\right) $
.  Having computed the kernel, one numerically inverts the statistical
model (eq. [\ref{M1}]) in terms of the observed $f_{\mathcal{A},\mathcal{T%
}}$ to determine the desired mass function characterizing the population
producing the microlenses.  With kinematically well-characterized
populations -- e.g., those (presumably) of the Galactic halo and LMC -- this
procedure appears to be a reasonable one for estimating the mass function of
the lensing population.

\section{STATISTICAL MODEL}

\subsection{Derivation}

The detailed statistical relation can be derived as follows.  Starting from
the marginalization tautology
\begin{equation}
f_{\mathcal{A},\mathcal{T}}\left( \mathcal{A},\mathcal{T}\right) =\int
\!\int \!\int f_{\mathcal{A},\mathcal{T},D_{L},D_{S},M}\left( \mathcal{A},
\mathcal{T},D_{L},D_{S},M\right) \,dD_{L}\,dD_{S}\,dM\;,  \label{M2}
\end{equation}
one changes variables within the integrand, from $\mathcal{A}$ and $\mathcal{
T}$ to $b=r_{E}\, \beta \left( \mathcal{A}\right) $ and $V_{\perp }=r_{E}/
\mathcal{T}$:
\begin{equation}
f_{\mathcal{A},\mathcal{T}}\left( \mathcal{A},\mathcal{T}\right) =\int
\!\int \!\int f_{b}\left( r_{E}\,\beta \left( \mathcal{A}\right) \right)
 \,f_{V_{\perp },D_{L},D_{S},M}\left( \frac{r_{E}}{\mathcal{T}}
,D_{L},D_{S},M\right) \,\left\vert {\bf J}\right\vert
\,dD_{L}\,dD_{S}\,dM\;.  \label{M3}
\end{equation}
As shown in Appendix A, in any likely application the impact
parameter $b$ will be uniformly distributed independently of any other
parameters, so that its distribution $f_{b}$ may be factored away from that
of the other parameters and assumed to be a constant that can be absorbed
into normalization of the model.  The quantity $\left\vert {\bf J}
\right\vert $ is the Jacobian of the variable transformation:
\begin{equation}
\left\vert {\bf J}\right\vert =\left\vert 
\begin{array}{ll}
\partial b/\partial \mathcal{A} & \partial b/\partial \mathcal{T} \\ 
\partial V_{\perp }/\partial \mathcal{A} & \partial V_{\perp }/\partial 
\mathcal{T}
\end{array}
\right\vert =\frac{r_{E}^{2}}{\mathcal{T}^{2}}\left[ 2\left( \frac{\mathcal{A
}}{\sqrt{\mathcal{A}^{2}-1}}-1\right) \left( \mathcal{A}^{2}-1\right) ^{3}
\right] ^{-1/2}\;.  \label{J}
\end{equation}
Combining these results,
\begin{equation}
f_{\mathcal{A},\mathcal{T}}\left( \mathcal{A},\mathcal{T}\right) =C\frac{L_{
\mathcal{A}}\left( \mathcal{A}\right) }{\mathcal{T}^{2}}\int \!\int \!\int
r_{E}^{2} \,f_{V_{\perp },D_{L},D_{S},M}\left( \frac{r_{E}}{\mathcal{T}}
,D_{L},D_{S},M\right) \,dD_{L}\,dD_{S}\,dM\;,  \label{M4}
\end{equation}
where $C$ is a constant and all of the amplitude information is contained
within the likelihood function
\begin{equation}
L_{\mathcal{A}}\left( \mathcal{A}\right) =\left[ \left( \frac{\mathcal{A}}{
\sqrt{\mathcal{A}^{2}-1}}-1\right) \left( \mathcal{A}^{2}-1\right) ^{3}
\right] ^{-1/2}\;,  \label{Lalpha}
\end{equation}
which is independent of any other parameters in the problem.  Equation (\ref
{M4}) makes clear that $\mathcal{A}$ and $M$ are effectively statistically
independent, so the lensing amplitude carries no significant mass
information.  We can thus treat $\mathcal{A}$ as a nuisance parameter and
integrate it from the model, leaving one in which the only observable
information is that of timescale $\mathcal{T}$.  The integral of $L_{
\mathcal{A}}$ actually diverges as $\mathcal{A}\rightarrow 1$, but that
lower limit corresponds to unobservably faint microlenses: in any practical
observing program there will be a minimum observable amplitude, however
poorly defined, that serves to determine the value of the integral of
equation (\ref{M4}) over $\mathcal{A}$, and to partly convert the left-hand
side of that equation to one of observed microlenses.  To complete that
conversion, we multiply the model by the probability $ E \left( \mathcal{T}
\right) $ that a microlens of timescale $\mathcal{T}$ will actually be
observed; this may be estimated by the observer based on properties of the
observing process, so the (non-normalized) model becomes one corrected for
observational selection:
\begin{equation}
f_{\mathcal{T}}^{\left( O\right) }\left( \mathcal{T}\right) \propto \frac{
E\left( \mathcal{T}\right) }{\mathcal{T}^{2}}\int \!\int \!\int
r_{E}^{2} \,f_{V_{\perp },D_{L},D_{S},M}^{\left( P\right) }\left( \frac{
r_{E}}{\mathcal{T}},D_{L},D_{S},M\right) \,dD_{L}\,dD_{S}\,dM\;.  \label{M4b}
\end{equation}
The superscript $\left( O\right) $ on the observable $f_{\mathcal{T}}$
denotes a distribution of observed lenses only, while the $\left( P\right) $
on the integrand refers to the distribution among the entire population of
objects from which the observed lenses and sources are drawn.  

Put nearly into the standard form of equation \ref{M1}, we have
\begin{equation}
f_{\mathcal{T}}^{\left( O\right) }\left( \mathcal{T}\right) =C\int L_{
\mathcal{T}|M}\left( \mathcal{T}|M\right) \, f_{M}^{\left( P\right)
}\left( M\right) \,dM\;,  \label{Mfinal}
\end{equation}
where
\begin{equation}
L_{\mathcal{T}|M}\left( \mathcal{T}|M\right) =\frac{E\left( \mathcal{T}
\right) }{\mathcal{T}^{2}}\int \!\int r_{E}^{2}\, f_{V_{\perp
},D_{L},D_{S}|M}^{\left( P\right) }\left( \frac{r_{E}}{\mathcal{T}}
,D_{L},D_{S}|M\right) \,dD_{L}\,dD_{S}\,\;  \label{kernel}
\end{equation}
is the likelihood of observing a microlens of timescale $\mathcal{T}$
arising from a lensing object of mass $M$ (which enters via $r_{E}$; cf.
eq. [\ref{RE}]). This -- equations \ref{Mfinal} and \ref{kernel} -- is the
desired result, the deterministic connection between lens population mass
function and observed timescale distribution (the normalizing factor $C$ is
to be computed \textit{a posteriori} from normalization of $f_{\mathcal{T}
}^{\left( O\right) }\left( \mathcal{T}\right) $). 

\subsection{A Corroborating Example}

As an example loosely based on observations of microlenses toward the LMC,
we postulate a thick disk whose stellar spatial density varies along the
line-of-sight to the LMC as $\rho \left( D\right) \propto \exp \left(
-D/l\right) $, where $l=8$ kpc is the projected scale length.  From Appendix C
we then expect the linear distribution of stellar distances
to be $f_{D_{L}}\left( D_{L}\right) \propto D_{L}^{2}\exp \left( -D/l\right) 
$.  We also postulate a Maxwellian velocity distribution for $V_{\perp }$
of $200$ km s$^{-1}$ rms, and put all the source stars at $D_{S}=55$ kpc. 
Assuming statistical independence of $D_{L}$, $D_{S},$ $V_{\perp }$, and $M$
; and a  timescale observing efficiency of $E\left( \mathcal{T}\right) =
\mathcal{T}\left( 200-\mathcal{T}\right) $ for $0\leq \mathcal{T}\leq 200$
days and $0$ otherwise; we use equation \ref{kernel} to compute the integral
kernel shown in Figure 1.
\begin{figure}
\hskip1in\includegraphics*[scale=0.75]{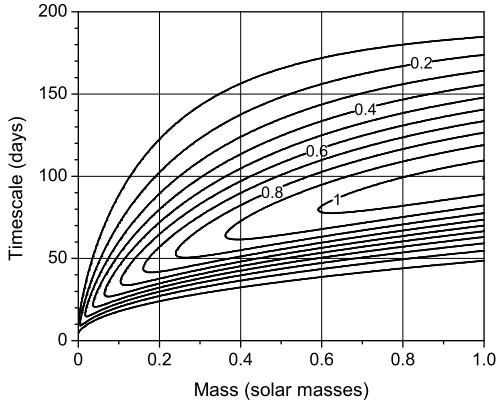}
\caption{Relative likelihood $L_{\mathcal{T}|M}\left( \mathcal{T}|M\right) $
(eq. [\ref{kernel}], arbitrarily scaled) that a compact object of
mass $M$ will produce an observable microlensing event of timescale $%
\mathcal{T}$, for a synthetic kinematic model and timescale observing
efficiency (see text).}
\end{figure}
With a presumed lensing population mass function uniform on $\left(
0,1\right) $ solar masses, this kernel yields (from eq. [\ref{Mfinal}])
the timescale distribution of observable microlenses shown in Figure 2.
\begin{figure}
\hskip1in\includegraphics*[scale=0.75]{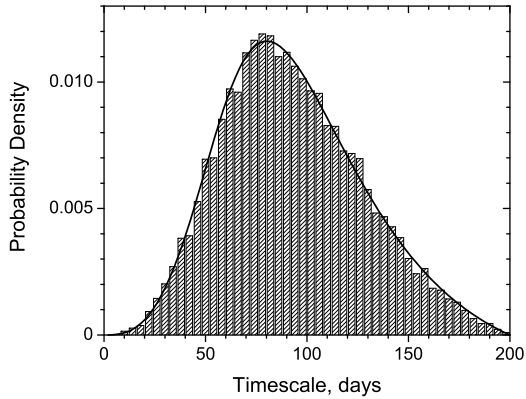}
\caption{Timescale distribution $f_{\mathcal{T}}\left( \mathcal{T}\right) $
for a hypothetical lensing population kinematic model and uniform mass function (see
text).  Solid curve: as predicted by the statistical model and computed
from the kernel function of Figure 1 and adopted mass function.  Histogram:
as simulated by Monte Carlo sampling from the presumed kinematic
distributions and mass function, without reference to the statistical model.}
\end{figure} 
Also shown in this figure are the corroborating results of one of many Monte
Carlo simulations with sample lensing object kinematics drawn directly from
the above distributions, without reference to the statistical model (the
simulations employed $b_{\max }=20$ A.U. (Appendix A) and $\mathcal{A}_{\min
}=1.3$ for observable lenses, neither of which affects the results; and
employs the same $E\left( \mathcal{T}\right) $ as used in the model).  The
excellent agreement between model predictions and an independent simulation
suggests an accurate model, fully corrected for timescale- and
amplitude-limited observational selection.

\subsection{Solution}

The model may be numerically inverted to determine the underlying population
mass function from an observed timescale distribution.  The simplest
inversion technique may be that of Richardson (1972) and Lucy (1974), in the
form proposed by Heacox (1997, eq. [A4]).  Suppressing the $\left(
P\right) $ and $\left( O\right) $ superscripts for clarity, this iterative
algorithm takes the form%
\begin{eqnarray}
\mathcal{L}^{\left( n\right) }\left( \mathcal{T},M\right) &=&\frac{L_{%
\mathcal{T}|M}\left( \mathcal{T}|M\right) }{\int L_{\mathcal{T}%
|M}(t|m)\,f_{M}^{\left( n\right) }\left( m\right) \,dm\,dt}\;,   \\
f_{M}^{\left( n+1\right) }\left( M\right) &=&f_{M}^{\left( n\right) }\left(
M\right) \int \frac{\mathcal{L}^{\left( n\right) }\left( t,M\right) \,f_{%
\mathcal{T}}\left( t\right) }{\int \mathcal{L}^{\left( n\right) }\left(
t,m\right) \,f_{M}^{\left( n\right) }\left( m\right) \,dm}\,dt\;.  \label{RL}
\end{eqnarray}%
One chooses an initial guess for $f_{M}^{\left( 1\right) }\left( M\right) $
(prudently chosen to be smooth on a wide range) and iterates until sensible
convergence. \ If the observed $f_{\mathcal{T}}^{\left( O\right) }$ is
actually derived from the presumed kinematic model, the method will almost
always converge to a reasonable approximation to the underlying mass
function. \ Note, however, that the statistical model is a smooth,
non-linear mapping from the space of mass functions into that of observed
timescale functions, for which we do not expect that every timescale distribution will have a legitimate mass function
pre-image, given the adopted kinematic model.  If either the timescale
function is inaccurate (due, e.g., to sampling variations) or the adopted
kinematic model is incorrect, the Richardson-Lucy method may fail to
converge when applied to this statistical model.

\section{APPLICATION TO GALACTIC HALO MICROLENSES}

\subsection{Data}

The lensing population responsible for microlenses observed against the
backdrop of the LMC probably represents the kinematically most homogeneous
set of objects to be analyzed for mass function with the statistical model
of the previous section.  The MACHO Project (Alcock et al. 2000) has
detected between 13 (Criteria A) and 17 (Criteria B) \textquotedblleft
simple\textquotedblright\ (point-source and point-mass) lensing events over
a 5.7-year period.  The timescale $ \left ( \mathcal(T) \right) $ distributions for these observations are
shown in Figure 3, together with their estimated observing efficiency $
E\left( \mathcal{T}\right) $ (note that the timescale $\hat{t}$ of Alcock et al. (2000) is $2\mathcal{T}$).
\begin{figure}
\hskip1in\includegraphics*[scale=0.85]{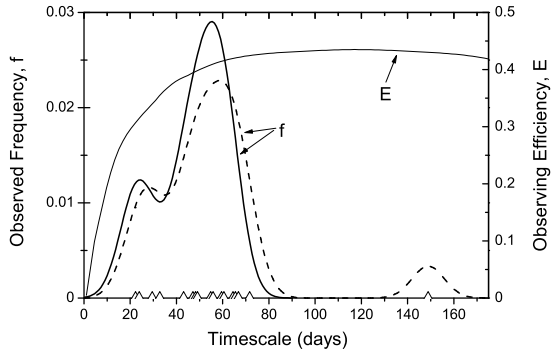}
\caption{Observed
microlens timescale distribution toward the LMC.  Solid, heavy curve:
Alcock et al.'s (2000) high-confidence events (\textquotedblleft Criteria
A\textquotedblright ); dashed curve, lower-confidence events
(\textquotedblleft Criteria B\textquotedblright ).  Light curve: estimated
timescale observability (a transcription of Alcock et al.'s (2002) 5-year
Criteria A curve of their Figure 5).  The symbols at the bottom denote the
timescales for individual Criteria B events (those of Criteria A are an
approximate subset).}
\end{figure}

Thetimescale frequency functions as shown are Gaussian kernel estimators
applied to Alcock et al's (2000) individual events (their Table 7), with a
smoothing kernel standard deviation set as large as possible ($7$ days)
consistent with the very low estimated probability of detecting events with
timescales less than $\sim 2$ days.  The multi-modality of these
distributions is quite possibly an artifact of small sample statistics, but
the narrow range of observed timescales probably is not (the matter is discussed further in \S 4).

\subsection{Kinematic Model}

There appears to be no unambiguous kinematic model of the dark halo with which these data may
be used in application of the statistical model.  While there is some dynamical evidence for the dark halo
mass distribution, there are few observational clues to the required velocity field even in the luminous halo.  
The root of the
difficulty is that the velocity of interest here -- that on the plane of the sky -- is unobservable
beyond the solar vicinity and cannot be reasonably inferred from equilibrium dynamical models based on
the collisionless Boltzmann equation (Binney \& Tremaine 1987).  Beyond the solar vicinity, 
such velocities are largely transverse to the Galactic center and thus reflect orbital angular momentum 
as much as they do orbital energy.  But angular momentum is essentially a free parameter in static equilibrium models, 
so that a whole sequence of models of different orbital angular momentum content may be fit to observed
spatial densities and radial velocity distributions (Heggie \& Hut 2003), and there is thus no effective
theoretical or observational constraint on halo transverse velocity distributions. 
Some evolutionary models of the luminous halo (e.g., Sommer-Larson et al. 1997) are consistent with reasonable levels of velocity anisotropy that, in conjunction with the observed radial velocity dispersion of distant halo stars, suggest a
large ($ > 100 $ km s$^{-1}$) transverse velocity dispersion throughout the halo. But it is not at all clear
that such speculative dynamical models of the luminous halo can be confidently applied to the dark halo.   

One is thus reduced to examining the compatibility of presumed dark halo
kinematic models with the observed microlens timescale distribution.  The
initial choice of kinematic model has here been made so that all parameters
are statistically independent (the alternative will be considered shortly)
and with the narrow range of observed timescales in mind, so that the ranges
of distances and velocities in the kinematic model have been kept as small
as is reasonably consistent with a spheroidal, pressure-supported halo.  Thus,
the ``NFW Law" of dark halo spatial mass
density, $\rho \left( R\right) \propto R^{-1}\left( a+R\right) ^{-2}$
(Navarro, Frank \& White 1997), is chosen\ as the basis of the MACHO
distance distribution, where $R$ is the galactocentric distance and $a=5$
kpc is the core radius. The resulting solarcentric distance distribution is
then (from Appendix C):
\begin{equation}
f_{D_{L}}\left( D_{L}\right) \propto D_{L}^{2}\left\{ R\left( D_{L}\right) 
\left[ a+R\left( D_{L}\right) \right] ^{2}\right\} ^{-1}\;,  \label{DL}
\end{equation}
where $R\left( D\right) =\left[ D^{2}+R_{\odot }^{2}-2DR_{\odot }^{2}\cos
b_{LMC}\cos l_{LMC}\right] ^{
{1/2}}$ 
with $R_{\odot }=8$ kpc presumed.  LMC source star distances are taken
to all be $55$ kpc (the more realistic assumption of a depth to the LMC
does not change the result).  The transverse velocity $V_{\perp }$ of the
lens relative to the source, as seen from our location, includes the motion
of the Sun relative to the LMC, and of the non-rotating halo rest frame
relative to the Sun.  The first of these is computed from the observed LMC
proper motions of $\mu _{\alpha ,\delta }=1.20,0.26$ micro-arcseconds per
year (Jones, Klemola \& Lin 1994) and an LMC\ distance of $55$ kpc, the
second from a presumed rotation rate of the Sun relative to the Halo of $200$
km s$^{-1}$.  The resulting mean velocity of the halo across the
line-of-sight to the LMC is thus $\left\langle V\right\rangle =215$ km s$
^{-1}$.  Relative to this mean frame we conservatively adopt a halo
velocity dispersion on the plane of the sky of $120$ km s$^{-1}$ (Binney \&
Merrifield 1998, Scheffler \& Els\"{a}sser 1988, Sommer-Larsen et al. 1997)
and realize this with low angular momentum orbits characterized by a dark
halo internal Gaussian random velocity distribution of mean $0$.  
With these and the presumption of no significant LMC
internal velocities, we compute (by the method of \S 3 of Appendix B)
the distribution of $V_{\perp }$ shown in Figure 4. 
\begin{figure}
\hskip1in\includegraphics*[scale=0.85]{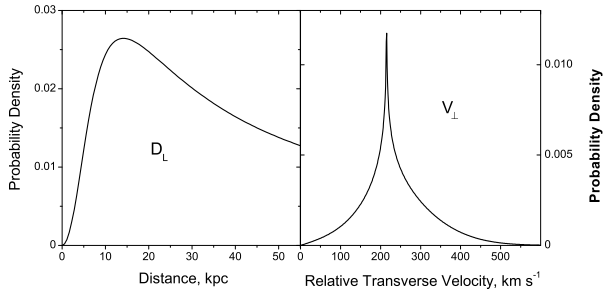}
\caption{Distributions of lens solarcentric distances $D_{L}$
(left panel) and relative transverse velocities $V_{\perp }$ (right panel) in
the assumed dark halo kinematic model (see text).}
\end{figure}

\subsection{Mass Function}

The integral kernel $L_{\mathcal{T}|M}\left( \mathcal{T}|M\right) $ computed
for these distributions (and assuming the $E\left( \mathcal{T}\right) $
estimated by Alcock et al. 2000 and illustrated in Figure 3) is similar in
appearance to that of the example shown in Figure 1. In application to
presumptive lens populations of a single mass (i.e., $f_{M}\left( M\right) $
as a delta function centered on the chosen mass), this model gives rise to
the sample timescale distributions shown in Figure 5.
\begin{figure}
\hskip1in\includegraphics*[scale=0.75]{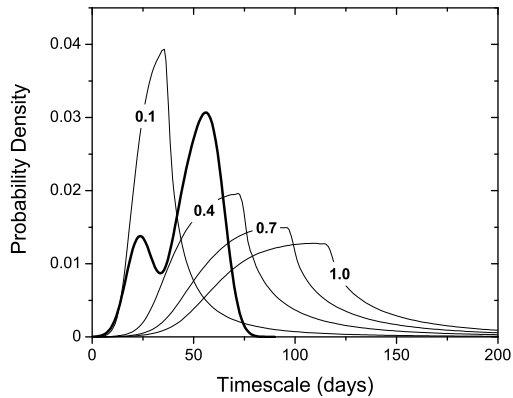}
\caption{Microlens time scale distributions (light
curves) predicted for single-mass lensing populations from a spheroidal,
pressure-supported, low angular momentum kinematic model of the halo (see
text).  The distributions are labeled with the stellar mass (solar masses)
that produced them.  The heavy curve is the adopted observed timescale
distribution (Criteria A of Alcock et al. 2000).}
\end{figure}
  
It is apparent that no non-pathological lens mass function can
adequately reproduce the observed timescale distribution, given this
kinematic model (the Richardson-Lucy interactive inversion method of \S 2.3,
applied to this model and Alcock et al.'s [2000] Criteria A observed timescale function, does not converge). 
The difficulty can be ascribed to the wide range of lens distances and
velocities in the kinematic model, which lead to similarly wide ranges of
observed timescales for given lens mass.  

The obvious alternative kinematic models do little, if anything, to rectify
this discrepancy.  Isothermal models, both with and without mass
segregation, lead to different kernels $L_{\mathcal{T}|M}\left( \mathcal{T}
|M\right) $ but similarly widely distributed timescale distributions
irrespective of lens mass. A choice of different dark halo mass distribution
(e.g., $\rho \left( R\right) \propto \left( a^{2}+R^{2}\right) ^{-1}$ or $
\left( a+R\right) ^{-2}$; Binney \& Merrifield 1998, Alcock et al. 1996,
1997) will typically exacerbate the problem, as will the inclusion of LMC
internal velocity dispersion in the computation of $V_{\perp }$.  The
timescale distribution widths may only be adequately decreased in a
spheroidal halo by (1) postulating a halo velocity field with a very small
transverse velocity dispersion, probably less than $\sim 50$ km s$^{-1}$,
while continuing to ignore the LMC internal motions; and/or (2) greatly
decreasing the spatial range encompassed by the lensing populations, as in a
thin shell.  Neither of these seems dynamically plausible.  Additional
caveats, based on the small sample size and poor numerical conditioning of
the statistical model, are discussed in \S 4.  For the moment, the
microlens timescale distribution as observed appears to be inconsistent with
a standard, spheroidal, pressure-supported model of the dark halo.

\subsection{Thick Disk Model}

As one possible alternative to the dark halo microlensing model we consider 
a rotating, low-dispersion, thick
disk as the site of microlensing objects.  To be consistent with the
roughly constant rotation velocity in the outer disk, we adopt a spatial
mass density that varies with galactocentric distance $r$ in the disk, and
height $z$ above the disk, of $\rho \left( r,h\right) \propto r^{-1}\exp
\left( -z/h\right) $. \ Following the argument of Appendix C,
the distribution of solarcentric distances will be%
\begin{equation}
f_{D_{L}}\left( D_{L}\right) \propto D_{L}^{2}\, r\left( D_{L}\right)
^{-1}\exp \left[ -\frac{D_{L}\sin \left\vert b_{LMC}\right\vert }{h}\right]
\;,  \label{td1}
\end{equation}
where $r\left( D\right) =\left[ D^{2}\cos ^{2}b_{LMC}+R_{\odot
}^{2}-2DR_{\odot }\cos b_{LMC}\cos l_{LMC}\right] ^{1/2}$ and we again
choose $R_{\odot }=8$ kpc.  To keep the timescale dispersions small we
choose $h=2$ kpc in a co-rotating thick disk of $30$ km s$^{-1}$ velocity
dispersion.  Both of these are at the low end of what would be expected,
both observationally and dynamically; but they seem possible.  We further
assume the statistical independence of all kinematic parameters, a constant
source distance of $55$ kpc, and an internal velocity dispersion of $30$ km s
$^{-1}$ for the LMC (but no rotation).  Then the integral kernel $L_{
\mathcal{T}|M}\left( \mathcal{T}|M\right) $ corresponding to this kinematic
model produces the single-mass timescale functions of Figure 6, from which
it is apparent that a distribution of lens masses encompassing roughly $0.1$
to $0.7$ solar masses may suffice to explain the MACHO observed timescale
distribution.
\begin{figure}
\hskip1in\includegraphics*[scale=0.75]{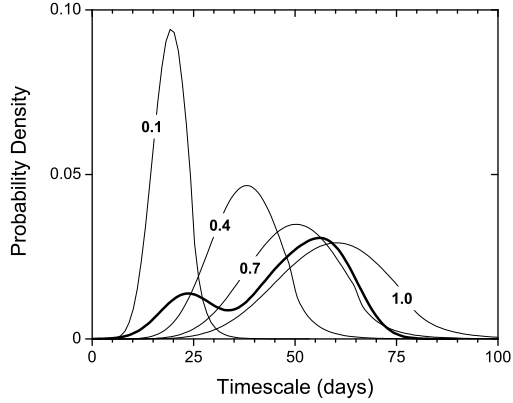}
\caption{As with Figure
5, but for a rotating, low velocity dispersion, thick disk population (see
text).  The thin curves are labeled with the lens mass (solar masses); the
thick curve is the MACHO observed timescale distribution (Alcock et al.
2000).}
\end{figure}  
In this case the Richardson-Lucy
inversion procedure of \S 2.3 converges to the results shown in Figure 7.
\begin{figure}
\hskip1in\includegraphics*[scale=1.0]{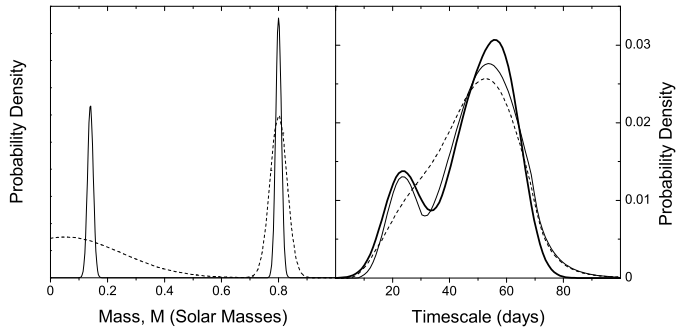}
\caption{Thick disk mass functions approximate solutions. \ Left panel:
mass functions (arbitrarily scaled) inferred from the Richardson-Lucy
algorithm applied to the statistical model derived from an adopted thick
disk kinematic model (see text) and observed timescale function (solid curve
of Figure 3).  Solid curve: 200 interactions beginning from a uniform mass
distribution.  Dashed curve: 100 interactions.  Right panel: the timescale
distributions (light curves) corresponding to the mass functions in the left
panel, as computed from the statistical model (eq. [\ref{Mfinal}]
and \protect\ref{kernel}). The heavy curve is the observed timescale
distribution.}
\end{figure} 
For this postulated thick disk there is no evidence for lensing objects
exceeding one solar mass, and some evidence that a significant fraction of
the lenses have low- or sub-stellar masses.

\section{SUMMARY AND CONCLUSIONS}

The statistical model of equations \ref{Mfinal} and \ref{kernel} is the
complete deterministic connection between the lens population mass function
and the distribution of timescales among observed microlenses.  It is fully
corrected for observational selection based on timescale and maximum
amplitude observability, providing these are mutually independent (which
will usually be the case).  The correction for selection on the basis of
amplitude arises from a basic model (eq. [\ref{M4}]) that includes the
amplitude as an observable, from which it is removed by integration over, in
effect, observable amplitudes only.  The result (easily proven by
construction of a basic model for timescales only) is the inclusion of an
additional factor of $r_{E}$ in the integrand: the observability of a
microlensing event, in terms only of amplitude, is proportional to the
Einstein radius of the lens, hence (approximately) to $M^{
{1/2}}$.

Appendix A provides a formal derivation of event impact parameter
distributions: for all practical purposes these are uniform on a range
exceeding that likely to be encountered in amplitude-limited observations. 
These results lead to a formal distribution of event amplitudes (eq. 
[\ref{Lalpha}]), which is independent of kinematic model parameters and of lens
mass; so that microlens mass information must come only from event
timescales.  These are both familiar results, here rigorously derived.

The statistical model is limited in applicability by its relatively poor
numerical conditioning, reflective of the small contribution of lens mass to
the event timescale.  The formal concept
of condition number is not strictly applicable to such non-linear models, but the sensitivity of the
observed timescale distribution to the mass function can be crudely
estimated as the change in mean timescale (as computed from the model) for a
given change in scale of a mass delta function.  For the example of \S 2.2,
this sensitivity is about $\Delta \left\langle T\right\rangle \approx 7$
days per $\Delta M=0.1$ solar masses over the range of $\left( 0.1,1\right) $
solar masses.  While this number depends upon the kinematic model adopted,
it is probably typical of most applications to widely distributed kinematics
and suggests a relatively poorly conditioned model from which mass function
information will be difficult to extract.  

The poor numerical conditioning of the model is reflected in its smoothing
effect in translating mass functions into timescale distributions.  This is
apparent from the thick disk model result of \S 3.4, where nearly
delta-function mass distributions serve to reproduce the smooth, bimodal
observed timescale distribution.  It is for this reason that no realistic
mass function can reproduce the substantial timescale structure in Alcock et
al.'s (2000) Criteria B sample (dashed curve in Figure 5): the isolated
event near 150 days, if real, is a true outlier, perhaps arising from a
separate lens population.

For microlensing toward the LMC, the conditioning difficulty is exacerbated
by uncertainties in the observed timescale distribution due to small sample
statistics.  Re-sampling of the presumed timescale distribution (Criteria A
of Alcock et al. 2000; heavy curve in Figure 5) demonstrates that a wide
range of timescale distributions is likely to arise from such a small
sample.  A significant fraction of 13-object distributions drawn from,
e.g., a Gaussian of mean $50$ days and standard deviation $15$ days (roughly
those of Criteria A events) show bi-modality similar to that of the Criteria
A distribution, so that the observed bi-modality may well be a spurious
consequence of small-number statistics.  

It thus seems unlikely that the current microlensing observations toward the
LMC could meaningfully constrain the (presumably) underlying dark halo mass
function.  Even such summary statistics as the mean mass are in some doubt:
since $M\propto V_{\perp }^{2}$ for given event timescale, and the mean of $
V_{\perp }$ depends as much upon the poorly known LMC proper motions as it
does on the solar galactocentric velocity, the canonical value of $
\left\langle M\right\rangle \approx 0.5$ solar mass (e.g., Alcock et al.
2000) may well be incorrect by a substantial factor.

But the narrow range of observed timescales appears to be real: re-sampling
experiments from this and similar distributions suggest little likelihood
that the true distribution is sufficiently broader than that observed as to
be compatible with kinematic distributions as broadly distributed as those
expected of a spheroidal, pressure-supported population of modest mean
orbital angular momentum.  It seems likely that the population producing
the microlenses observed toward the LMC must come from something other than
a \textquotedblleft standard\textquotedblright\ halo, either dark or
luminous.  The results of \S 3.4 suggest that a thick disk origin is one
possibility; no doubt there are others.

\newpage
\appendix

\section{IMPACT PARAMETER DISTRIBUTION}

We take as the most likely geometric model one in which, at the commencement
of the survey, the lensing candidates are uniformly distributed on the plane
of the sky in the vicinity of the source star being observed. The geometry
of the encounter is described entirely by the original distance between
lensing object and source star, $r$, and the direction $\xi $ of motion
relative to the radius vector from star to lens; both measured in the lens
plane. The impact parameter corresponding to $\left( r,\xi \right) $ will
then be $b=r\sin \xi $; $r$ will be distributed as $f_{r}\left( r\right)
=2r/r_{\max }^{2}$ where $r_{\max }$ is the limiting radius of the search
area in the lens plane; and, by virtue of the uniform distribution of lens
locations, we can safely presume $\xi $ to be uniformly distributed on $
\left( 0,\pi /2\right) $.  By a useful theorem of cumulative distributions
(e.g., \S 7.26 of Kendall, Stuart and Ord 1987), the pdf of $b$ is given by%
\begin{equation}
f_{b}\left( b\right) =\frac{d}{db}\int_{\Omega }\int f_{r}\left(
r\right) f_{\xi }\left( x\right) \,d\xi \,dr\;,
\end{equation}
where $\Omega =\left\{ \left( r,\xi \right) |r\sin \xi <b\right\} $ is the
set of all lens coordinates for which the impact parameter is less than $b$.
\ Thus:
\begin{eqnarray}
f_{b}\left( b\right) &=&\frac{d}{db}\left[ \int_{0}^{b}\int_{0}^{\pi
/2}+\int_{b}^{r_{\max }}\int_{0}^{\arcsin \left( b/r\right) }\right] \left( 
\frac{4r}{\pi r_{\max }^{2}}\right) \,d\xi \,dr\;,   \\
&=&\frac{4}{\pi r_{\max }}\left[ 1-\left( \frac{b}{r_{\max }}\right) ^{2}
\right] ^{{1/2}}\;,  \label{fb2}
\end{eqnarray}
for $0\leq b\leq r_{\max }$ and $0$ otherwise.  This formula has been
verified by Monte Carlo simulations based on the above geometric model.  In
most applications the largest impact parameter corresponding to observable
amplitudes will be smaller than $r_{\max }$ by several orders of magnitude
(about $6-10$ for observable halo microlenses) and $f_{b}\left( b\right) $
may safely be taken to be a constant independent of all other parameters. 
The above result, with the assumption $b\ll r_{\max }$, leads to the
amplitude likelihood of equation \ref{Lalpha}.  From that relation the
distribution of scaled impact parameter $\beta $ is easily seen to be
\begin{equation}
f_{\beta }\left( \beta \right) \propto L_{\mathcal{A}}\left( \mathcal{A}
\left( \beta \right) \right) \left\vert \frac{d\mathcal{A}}{d\beta }
\right\vert =1\;.  \label{fbeta}
\end{equation}
The distributions of observable microlens impact parameters, scaled or not,
are thus expected to be uniform and independent of distributions of lens
distances, velocities, or masses.

\section{VELOCITY DISTRIBUTIONS}

The velocity $V_{\perp }$ is the relative velocity of lens and source star,
projected onto the plane of the sky, as seen from our location.  It can
usually be computed as the magnitude of the vector sum of two or more
projections of velocity components onto the plane of the sky.  For many
applications, the following results will suffice to compute the distribution
of this quantity, from a kinematic model of the three-dimensional velocity
distributions of lenses and source stars.

\subsection{Projection onto Plane of Sky}

An isotropic space velocity $V$ will have a magnitude projected onto the plane of the
sky, $V_{pos}$, given by the simple projection relation encountered in,
e.g., analysis of binary star orbits (Chandresakhar and M\"{u}nch 1950):
\begin{equation}
f_{V_{pos}}\left( V_{pos}\right) =V_{pos}\int\limits_{V_{pos}}^{\infty }
\frac{f_{V}\left( V\right) }{V\sqrt{V^{2}-V_{pos}^{2}}}\,dV\;.  \label{pos}
\end{equation}

\subsection{Combination of Random Velocities}

Two statistically independent, mutually orthogonal, random velocity
components $\left( V_{x},V_{y}\right) $ on the plane of the sky; have a
combined magnitude $V=\left [ V_{x}^{2}+V_{y}^{2} \right ]^{1/2} $ at a position angle $%
\theta =\arccos \left( V_{x}/V\right) $, whose joint distribution is given
by the simple variable transformation:
\begin{equation}
f_{V,\theta }\left( V,\theta \right) =V\, f_{V_{x}}\left( V\sin \theta
\right) \, f_{V_{y}}\left( V\cos \theta \right) \;  \label{Vtheta}
\end{equation}
($V$ is the Jacobian).  The marginal distribution of velocity magnitude
alone is given by the integral of this expression over $\theta $ or,
changing variables to $V_{x}=V\sin \theta $,
\begin{equation}
f_{V}\left( V\right) =V\int\limits_{0}^{V}\frac{f_{V_{x}}\left(
V_{x}\right) \, f_{y}\left( \sqrt{V^{2}-V_{x}^{2}}\right) }{\sqrt{
V^{2}-V_{x}^{2}}}\,dV_{x}\;.  \label{twoV}
\end{equation}

\subsection{Combination of Random \& Streaming Velocities}

The distribution of magnitudes of the combination of a random velocity $
V_{R} $ with a streaming velocity $V_{S}$ (e.g., that arising from the LMC
proper motions), both on the plane of the sky, may be computed as follows. 
Let $f_{V_{R},\theta }\left( V_{R},\theta \right) $ be the joint
distribution of random velocity magnitudes and directions (as measured from
the streaming velocity direction), as computed by equation \ref{Vtheta} or,
e.g., as a Gaussian velocity whose mean and/or standard deviation is a
function of direction.  Then, following the same procedure as with equation 
\ref{twoV}, the resulting velocity magnitudes on the plane of the sky will
be distributed as
\begin{equation}
f_{V_{\perp }}\left( V_{\perp }\right) =V_{\perp }\int\limits_{0}^{\pi }
\frac{f_{V_{R},\theta }\left( V_{R}\left( V_{\perp },\phi \right) ,\theta
\left( V_{\perp },\phi \right) \right) }{V_{R}\left( V_{\perp },\phi \right) 
}d\phi \;,  \label{S1}
\end{equation}
where 
\begin{eqnarray*}
V_{R}\left( V_{\perp },\phi \right) &=&\left [V_{S}^{2}+V_{\perp
}^{2}-2V_{S}V_{\perp }\cos \phi \right ] ^{1/2}\;, \\
\sin \left( \theta \left( V_{\perp },\phi \right) \right) &=&\frac{V_{\perp
}\sin \phi }{V_{R}\left( V_{\perp },\phi \right) }\;,
\end{eqnarray*}
and $V_{\perp }/V_{R}\left( V_{\perp },\phi \right) $ is the Jacobian of the
variable transformation from $\left( V_{R},\theta \right) $ to $\left(
V_{\perp },\phi \right) $. \ If the random component is isotropic, this
reduces to the relatively simple expression
\begin{equation}
f_{V_{\perp }}\left( V_{\perp }\right) =\frac{2V_{\perp }}{\pi }
\int\limits_{\left\vert V_{S}-V_{\perp }\right\vert }^{V_{S}+V_{\perp }}
\frac{f_{V_{R}}\left( V_{R}\right) }{\sqrt{\left( 2V_{S}V_{\perp }\right)
^{2}-\left( V_{S}^{2}+V_{\perp }^{2}-V_{R}^{2}\right) ^{2}}}\,dV_{R}\;.
\label{S2}
\end{equation}
An example may be seen in Figure 4.  This rather unusual function has been
verified by Monte Carlo simulations independent of the statistical reasoning
underlying this derivation.  If $V_{S}\rightarrow 0$ this expression
reduces to $f_{V_{\perp }}=f_{V_{R}}$, as it must.

\section{DISTANCE DISTRIBUTION}

If the spatial number density of stars varies with galactocentric distance $R
$ as $\rho \left( R\right) $, and the distribution is radially symmetric
about the point from which $R$ is measured, the linear density (number per
unit change in $R$) is easily seen to be $f_{R}\left( R\right) \propto
R^{2}\rho \left( R\right) $.  But viewed from off-center, or if the spatial
distribution is not radially symmetric, the matter is complicated by the
dependence of the density upon the viewing direction.  In the case of
microlenses viewed against the LMC, the field of view is considerable and it
is not obvious how the spatial number density will translate into a linear
density in solarcentric distance $D$ when viewed from our off-center
position. \ Formally the solution is straightforward: by the usual change of
variables,
\begin{equation}
f_{D,b,l}\left( D,b,l\right) \propto R^{2}\, \rho \left( R\right)
\left\vert \frac{\partial \left( R,\phi ,\theta \right) }{\partial \left(
D,b,l\right) }\right\vert \;,  \label{JD}
\end{equation}
where $R=\left[ R_{\odot }^{2}+D^{2}-2R_{\odot }D\cos b\cos l\right] ^{
{1/2}}$ and $\left( \phi ,\theta \right) $ are the galactocentric angular
coordinates for which we presume connecting relations with the solarcentric
position (e.g, $\sin \theta =D\sin b/R\left( D,b,l\right) $, etc.).  The
average value for the field of view is just the integral of this density
over the corresponding spreads in $b$ and $l$.  The Jacobian of this
relation is difficult to represent analytically, but is numerically trivial.
 As the field of view shrinks the spatial density becomes more uniform over
that field until, in the limit of a small field of view, the solarcentric
distance pdf becomes very nearly
\begin{equation}
f_{D,b,l}\left( D,b,l\right) \propto D^{2}\cdot \rho \left( R\left(
D,b,l\right) \right) \;,  \label{fD}
\end{equation}
where $\left( b,l\right) $ refer to the center of the field.  In numerical
experiments for the LMC direction, with a $\sim 10^{0}$ wide field and an
NFW spatial density profile (cf. \S 3.2), this approximation agrees with the
more exact result (eq. [\ref{JD}]) to within better than $1\%$, when
integrated over the field.

\clearpage

\end{document}